\documentclass[preprint,amsmath,amssymb,aps,prl,superscriptaddress,nobibnotes]{revtex4-2}

\usepackage{graphicx}
\usepackage{dcolumn}
\usepackage{bm}
\usepackage{xcolor}

\newcommand{\parallelsum}{\mathbin{\!/\mkern-5mu/\!}}

\begin{document}

	\title{From orbital to paramagnetic pair breaking in layered superconductor 2H-NbS$_2$}

\author{D.~Pizzirani}
\thanks{These authors contributed equally.}
\affiliation{High Field Magnet Laboratory (HFML-EMFL), Radboud University, Toernooiveld 7, Nijmegen 6525 ED, The Netherlands.}
\affiliation{Radboud University, Institute for Molecules and Materials, Nijmegen 6525 AJ, The Netherlands.}

\author{T.~Ottenbros}
\thanks{These authors contributed equally.}
\affiliation{High Field Magnet Laboratory (HFML-EMFL), Radboud University, Toernooiveld 7, Nijmegen 6525 ED, The Netherlands.}
\affiliation{Radboud University, Institute for Molecules and Materials, Nijmegen 6525 AJ, The Netherlands.}

\author{M.~van Rijssel}
\affiliation{High Field Magnet Laboratory (HFML-EMFL), Radboud University, Toernooiveld 7, Nijmegen 6525 ED, The Netherlands.}
\affiliation{Radboud University, Institute for Molecules and Materials, Nijmegen 6525 AJ, The Netherlands.}

\author{O.~Zheliuk}
\affiliation{High Field Magnet Laboratory (HFML-EMFL), Radboud University, Toernooiveld 7, Nijmegen 6525 ED, The Netherlands.}
\affiliation{Radboud University, Institute for Molecules and Materials, Nijmegen 6525 AJ, The Netherlands.}

\author{Y.~Kreminska}
\affiliation{Device Physics of Complex Materials, Zernike Institute for Advanced Materials, University of Groningen, 9747 AG Groningen, The Netherlands.}

\author{M.~Rösner}
\affiliation{Radboud University, Institute for Molecules and Materials, Nijmegen 6525 AJ, The Netherlands.}

\author{J.~F.~Linnartz}
\affiliation{High Field Magnet Laboratory (HFML-EMFL), Radboud University, Toernooiveld 7, Nijmegen 6525 ED, The Netherlands.}
\affiliation{Radboud University, Institute for Molecules and Materials, Nijmegen 6525 AJ, The Netherlands.}

\author{A.~de~Visser}
\affiliation{Van der Waals-Zeeman Institute, University of Amsterdam, Science Park 904, 1098 XH Amsterdam, The Netherlands.}

\author{N.~E.~Hussey}
\affiliation{High Field Magnet Laboratory (HFML-EMFL), Radboud University, Toernooiveld 7, Nijmegen 6525 ED, The Netherlands.}
\affiliation{Radboud University, Institute for Molecules and Materials, Nijmegen 6525 AJ, The Netherlands.}
\affiliation{H.H. Wills Physics Laboratory, University of Bristol, Bristol BS8 1TL, United Kingdom.}

\author{J.~Ye}
\affiliation{Device Physics of Complex Materials, Zernike Institute for Advanced Materials, University of Groningen, 9747 AG Groningen, The Netherlands.}

\author{S.~Wiedmann}
\affiliation{High Field Magnet Laboratory (HFML-EMFL), Radboud University, Toernooiveld 7, Nijmegen 6525 ED, The Netherlands.}
\affiliation{Radboud University, Institute for Molecules and Materials, Nijmegen 6525 AJ, The Netherlands.}

\author{M.~R.~van~Delft}
\email{maarten.vandelft@ru.nl}
\affiliation{High Field Magnet Laboratory (HFML-EMFL), Radboud University, Toernooiveld 7, Nijmegen 6525 ED, The Netherlands.}
\affiliation{Radboud University, Institute for Molecules and Materials, Nijmegen 6525 AJ, The Netherlands.}
	
	\date{\today}
	
	\begin{abstract}
 The superconducting transition metal dichalcogenides 2H‐NbSe$_2$ and 2H‐NbS$_2$ are intensively studied on account of their unique electronic properties such as Ising superconductivity, found in multi- and monolayers, with upper critical fields beyond the Pauli limit. Even in bulk crystals, there are reports of multiband superconductivity and exotic states, such as the Fulde-Ferrell-Larkin-Ovchinnikov phase. In this work, we investigate the superconducting properties of 2H‐NbS$_2$ through a detailed high-field mapping of the phase diagram by means of magnetotransport and magnetostriction experiments. We compare the phase diagram between bulk crystals and a 6~nm thick flake of 2H‐NbS$_2$ and find a drastically enhanced Maki parameter in the flake, signifying a change of the relevant pair breaking mechanism from orbital to paramagnetic pair breaking, which we attribute to an effect of enhanced spin-orbit coupling. 
	\end{abstract}
	
	\maketitle
\newpage
The family of transition metal dichalcogenides (TMDs) receives significant attention due to the tunability of its electronic ground states. TMDs, due to their Van der Waals bound layers, can be exfoliated towards the atomic limit to enabling exotic physics such as Ising superconductivity \cite{Zhou2016a,DeLaBarrera2018,Xing2017}. However, even in their bulk form, they exhibit interesting behavior, playing host to coexisting or competing ground states such as superconducting and charge density wave (CDW) phases \cite{Majumdar2020,Sanna2022,Moulding2020}. 

Within this family, 2H-NbS$_2$ stands out as the only superconducting compound that does not exhibit a CDW, despite indications that it may be close to a CDW transition \cite{Leroux2012a,Leroux2018,Bianco2019,Wen2020}. Because of this, studies of the superconducting state of 2H-NbS$_2$  (we drop the 2H in the following) compared to that of other family members such as NbSe$_2$ may hold clues as to the interplay between the CDW and superconductivity. NbS$_2$ has a critical temperature, $T_c$, of approximately 6~K \cite{Majumdar2020,Tissen2013,Kacmarcik2010,Lian2017} and has been reported to be a two-gap superconductor \cite{Guillamon2008,Kacmarcik2010,Kacmarcik2010a} and to exhibit orbital-selective two-dimensional superconductivity \cite{Bi2022} as well as a Fulde-Ferrell-Larkin-Ovchinnikov (FFLO) state \cite{Cho2021}.

In this work, we determine the magnetic field-temperature ($H$-$T$) phase diagram of bulk NbS$_2$ using magnetotransport and magnetostriction in magnetic fields up to 30~T, as well as that of a thin flake of NbS$_2$ using only magnetotransport. Motivated by previous reports, we look for signatures of exotic superconducting states, such as sharp enhancements of the critical field at low temperatures and particular field orientations. With our transport measurements, we extend previously reported phase diagrams \cite{Kacmarcik2010,Onabe1978,Cho2021} down to 0.3~K and investigate the dimensionality of the superconducting state in this layered type-II superconductor. Through a Werthamer-Helfand-Hohenberg (WHH) model description of the phase diagrams, we investigate the superconducting pair breaking mechanism, which we find to be primarily orbital in the bulk crystals, but paramagnetic in the thin flake. Importantly, the suppression of the orbital mechanism in the thin flake does not lead to an enhancement of the upper critical field $H_{c2}$. Rather, an enhanced Pauli paramagnetic pair breaking causes a reduced $H_{c2}$.

\section*{Methods}
Millimeter size single crystals of NbS$_2$ were purchased from HQ Graphene and cut into bars approximately 3$\times$0.7$\times$0.03~mm in size. Electrical contacts to these bulk crystals were made using silver paint (Dupont 4929). It was found that these contacts degrade over time, and had to be remade shortly before each experiment. 

Thin flakes of NbS$_2$ were exfoliated by the standard scotch tape method in an inert argon atmosphere inside a glovebox. The film was immediately protected by resist PMMA A4 and air exposure was minimized throughout the whole fabrication process. Electron beam lithography was used to expose the electrode regions, followed by a metallization process with a Pt/Au film of 5/40 nm thick. After liftoff and bonding, the sample was stored in vacuum prior to low temperature characterization. 

Electrical magnetotransport measurements were performed in superconducting magnets up to 13 or 16~T and in a Bitter magnet up to 30~T using standard lock-in techniques and an applied current up to 1~mA for bulk samples and 10~$\mu$A for the thin flake. 

Complementary measurements of the bulk thermodynamic properties were performed using a capacitive dilatometer technique \cite{Kuchler2017} attached to an in-situ rotation mechanism. Multiple single crystals were stacked together along the $c$-axis and a magnetic field up to 20~T was applied parallel to the planes. The magnetostriction curves were recorded with a capacitance bridge at a constant temperature using sweep rates of 1.0~T/min in order to minimize eddy current effects. The results of the magnetostriction experiments are discussed in the supplementary information.

\section*{Results}
\begin{figure}[h!]
	\centering
	\includegraphics[width=0.85\linewidth]{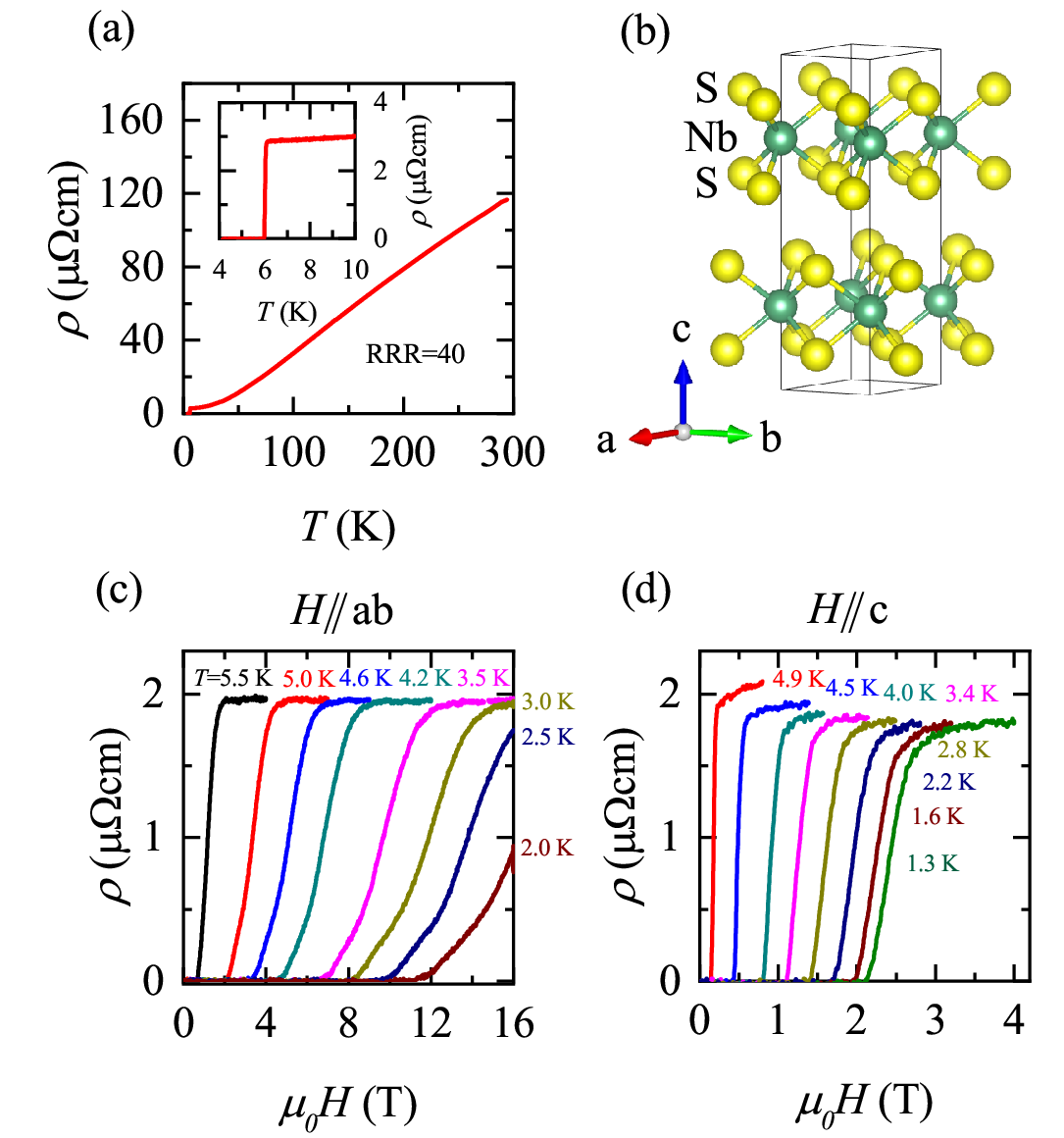}
	\caption{(a) Temperature dependence of the resistivity in bulk NbS$_2$. The inset highlights the range around the superconducting transition. (b) Crystal structure of NbS$_2$, with lattice parameters $a=b=3.31$~\AA~ and $c=11.89$~\AA~\cite{Jellinek1960}. Image generated with Vesta \cite{Momma2011}. (c),(d) Field-dependent resistivity at several temperatures, with the magnetic field applied (c) parallel to the planes ($H\parallelsum ab$) and (d) perpendicular to the planes ($H\parallelsum c$). }
	\label{fig1}
\end{figure}

In Fig.~\ref{fig1}(a), we show the temperature dependence of the resistivity of a typical bulk NbS$_2$ crystal. We can clearly identify a sharp superconducting transition (less than 0.15~K wide) around $T_c=6$~K and, as expected, no resistive signature of a CDW transition as occurs in NbSe$_2$. The residual resistivity ratio (RRR) of our bulk samples is approximately 40, indicating that our material is of comparable quality to that used in previous reports \cite{Wen2020,Huang2022b,Yan2019, Martino2021,Lian2017,Bag2018,Onabe1978}. 

Field-dependent data of the resistivity at different temperatures are plotted for magnetic fields parallel ($H\parallelsum ab$, $\theta=90^{\circ}$) and perpendicular ($H\parallelsum c$, $\theta=0^{\circ}$) to the planes of the crystal (see Fig.~\ref{fig1}(b)) in Figs.~\ref{fig1}(c) and (d), respectively. The main feature here is the increase in transition width with decreasing temperature that is especially pronounced in parallel field. Such behavior is typically seen in the case of two-dimensional (2D) superconductivity  \cite{Singleton2001,Kasahara2020a,Szabo2001,Vicent1980,Baidya2021}. The change in slope during the transition that can also be seen in the $B\parallelsum ab$ curves can be attributed to sample inhomogeneity. 

The angle dependence of the critical field, $H_{c2}$, can be used to identify the dimensionality of the superconducting state. Before going into that, however, it is important to consider the definition of $H_{c2}$. There is no general agreement within the literature about which point should be used, but a common procedure is to consider the field at which a certain percentage (e.g. 90\%) of the normal state resistance, $R_n$, is reached. An alternative method, especially useful when $R_n$ cannot be unambiguously determined due to a limited field range of normal state resistivity or a large magnetoresistance, is to look at the first or second derivatives of the signal. Minima or maxima in these derivatives can then be used as consistent points to locate $H_{c2}$. In supplementary Fig.~S1, we summarize the different definitions. We have employed all of these throughout our study, and found that they largely reproduce the same behavior. Within this paper, we use the 50\%$R_n$ criterion unless otherwise indicated.

\begin{figure}[h!]
	\centering
	\includegraphics[width=0.85\linewidth]{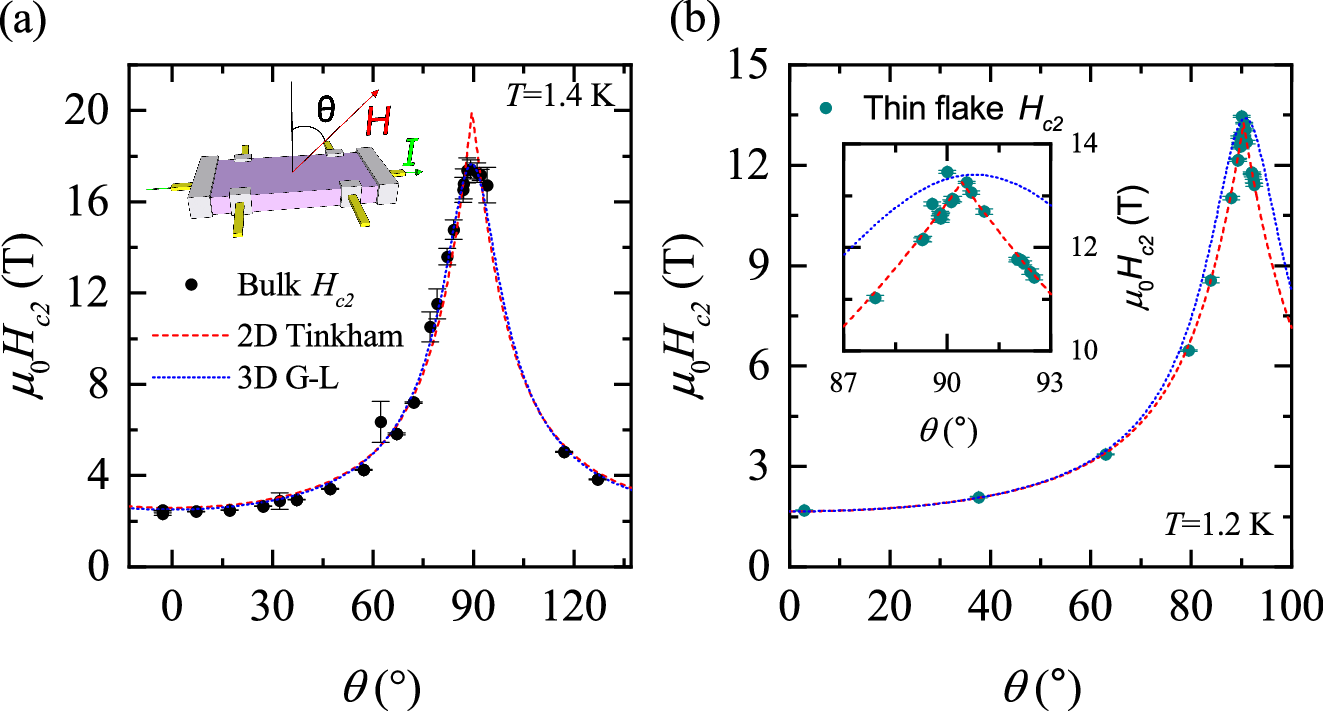}
	\caption{(a) Critical field of bulk NbS$_2$ (defined as the field where the resistance reaches 50\% of the normal state value) as a function of angle, at a temperature of 1.4~K. Dashed lines are fits to the two-dimensional Tinkham and three-dimensional Ginzburg-Landau models (see SI for the complete models). Inset: configuration of the sample showing the field orientation. 
		(b) Critical field of a 6~nm thick NbS$_2$ thin flake as function of angle at $T$=1.2~K, fit with the same models as (a). Inset: enlarged view of the region around 90$^{\circ}$.
	 }
	\label{fig2}
\end{figure}

In Fig.~\ref{fig2}(a) and (b), we show the angle dependence of $H_{c2}$ of a representative bulk NbS$_2$ sample and a 6~nm thin flake, respectively. We attempt to describe the data using the Tinkham model \cite{Tinkham1963} for 2D and the Ginzburg-Landau (G-L) model \cite{Tinkham1996} for three-dimensional (3D) superconductors (corresponding equations can be found in the supplementary information (SI)). As can be seen, 3D G-L provides a better description in the bulk case, particularly near $\theta=$90$^{\circ}$, suggesting that superconductivity is not confined to the individual layers, but a 3D state is formed instead. For the 6~nm thin flake, on the other hand, the 2D model clearly gives a better fit of the data, capturing the sharp cusp expected from the Tinkham model. This is somewhat surprising, as the sample thickness still exceeds the out-of-plane coherence length (of approximately 2~nm as we will discuss later) and so this flake should normally be considered bulk despite being four orders of magnitude thinner than our actual bulk crystals. 

\section*{Discussion}
In the following, we will analyze several aspects of the superconductivity in NbS$_2$. In particular, we discuss the question of dimensionality of the superconducting state and the manner of breaking of that state, for both bulk crystals and thin flakes.

\begin{figure}[h!]
	\centering
	\includegraphics[width=0.475\linewidth]{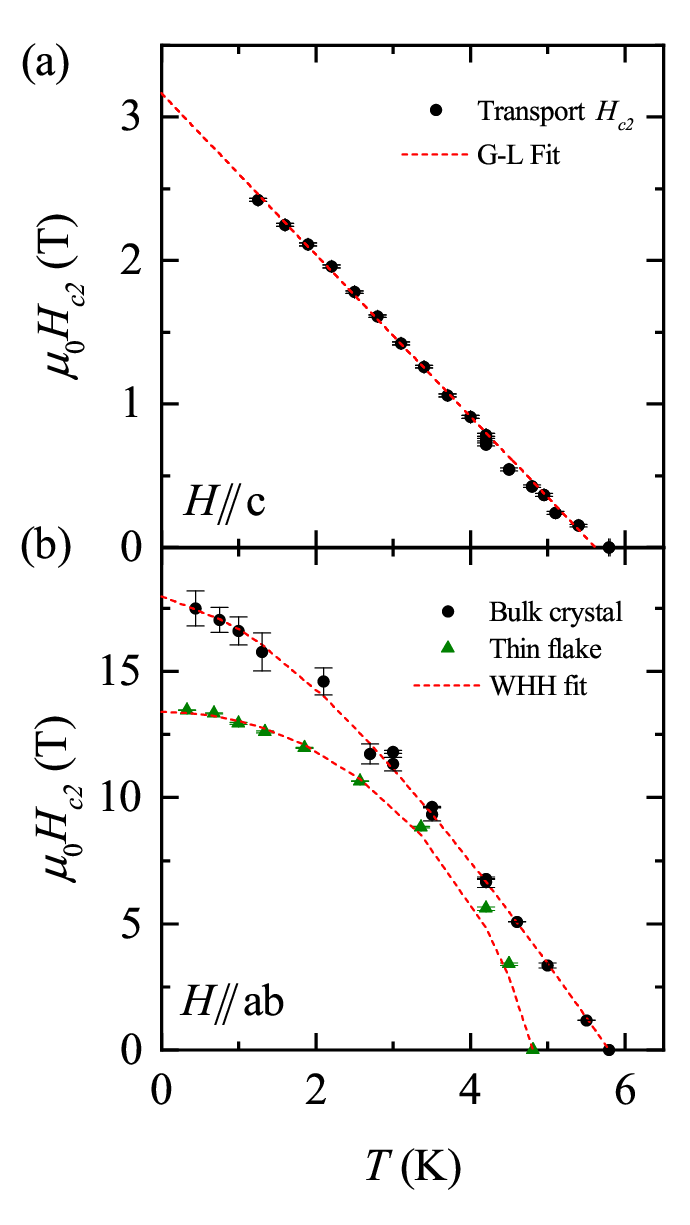}
	\caption{(a) $H$-$T$ phase diagram of bulk NbS$_2$ for $H\parallelsum c$. The dashed line is a linear fit using the G-L model for a two-dimensional superconductor under perpendicular field. (b) Comparison of the phase diagrams for bulk and thin flake NbS$_2$ (with $H\parallelsum ab$). The WHH fit for the thin flake data gives $\alpha_M=6\pm1$ and $\lambda_{SO}=1\pm0.3$. }
	\label{fig4}
\end{figure}

In Fig.~\ref{fig4}(a) and (b), we show the $H$-$T$ phase diagrams of bulk NbS$_2$ for $H\parallelsum c$ and $H\parallelsum ab$, respectively. For $H\parallelsum c$, the upper critical field depends linearly on temperature, matching the expectation for a 2D system in the Ginzburg-Landau model \cite{Sharma2018,Lu2015,Zeng2018}. In this case, the linearity is preserved for all definitions of $H_{c2}$ as well as $H_{irr}$. This could suggest that the superconducting state in bulk NbS$_2$ may have a certain 2D character, at least under the influence of an out-of-plane applied magnetic field. Alternatively, $\mu_0 H_{c2}(T)$ may also flatten off below our measured temperature range. In that case, a two-gap model can explain the data  \cite{Xing2017b}, which would be consistent with previous reports of two-gap superconductivity in NbS$_2$ \cite{Kacmarcik2010a,Guillamon2008}.

For $H\parallelsum ab$, however, the G-L model \cite{Tinkham1996}, which does not capture Pauli limiting effects, cannot accurately describe the data (see supplementary Fig. S8). Instead, both $H_{irr}$ and $H_{c2}$ can be well described by the Werthamer-Helfand-Hohenberg (WHH) model \cite{Werthamer1966} that considers both the Pauli paramagnetic and orbital effects, as shown in Fig.~\ref{fig4}(b) (details of the WHH model fitting can be found in the SI).

In general, the Cooper pairs in a superconductor may be broken either via the spin paramagnetic effect, where the magnetic field flips one spin of the spin-singlet Cooper pair to align both spins with the field direction; or via the orbital effect, where the Lorenz force exceeds the binding force between the electrons. Based on the WHH fitting, the spin paramagnetic effect is negligible in bulk NbS$_2$, as follows from the extracted value of the Maki parameter. We find that this parameter, which quantifies the relative strength of orbital and paramagnetic pair breaking, vanishes (i.e. $\alpha_M=\sqrt{2}\frac{H_{c2}^{orb}(0)}{H_{c2}^{pm}(0)}=0$, where $H_{c2}^{orb}$($H_{c2}^{pm}$) is the orbitally (paramagnetically) limited critical field). From this we conclude that orbital pair breaking is the dominant mechanism for breaking superconductivity in bulk NbS$_2$. This observation is inconsistent with the existence of an FFLO state in this material, which requires $\alpha_M>1.8$ \cite{Wosnitza2018,Gruenberg1966}. 

We now compare the $H$-$T$ phase diagrams of bulk and thin flake NbS$_2$. Clearly, both $T_c$ and $H_{c2}$ are reduced in the thin flake, but also the shape of the $\mu_0 H_{c2}(T)$ curve is notably different.  Applying the WHH model to the thin flake data, we find $\alpha_M=6\pm1$ and $\lambda_{SO}=1.0\pm0.3$, with the error bars estimated from the distribution of values found by fitting to different definitions of $H_{c2}$ as well as the experimental determination of $\left(\frac{dH_{c2}}{dT}\right)_{T=T_c}$ (see supplementary table S1). This is a dramatic change compared to bulk NbS$_2$ and indicates a strong influence of Pauli paramagnetic pair breaking in addition to orbital pair breaking. 

The key difference between our data and other reports \cite{VanWeerdenburg2023,Nam2018} of increasing Maki parameter with decreasing material thickness, is that we see a suppressed instead of increased $H_{c2}$, as could be understood from quenching of orbital pair breaking in a confined superconductor where screening currents cannot build up for $H\parallelsum ab$ \cite{VanWeerdenburg2023,Nam2018}. This is typically the case when a sample thickness of the order of $\xi_c$ or smaller is realized, as the orbital pair breaking effect should be absent when the thickness is smaller than the coherence length \cite{Matsuda2007}. An increase of the Maki parameter with suppressed $H_{c2}$ after thinning down a material has only been previously reported in FeSe flakes \cite{Farrar2020}, though the effect was not explained and even bulk FeSe is Pauli limited, so the circumstances are notably different.

In order to investigate the possibility of quenched orbital pair breaking, we determine $\xi$ for our samples. In general, $\xi$ can be extracted from the orbital critical fields \cite{Tinkham1996,Farrar2020}, giving $\xi_c$=1.9$\pm$0.2~nm and $\xi_{ab}$=10.2$\pm$0.5~nm for the bulk crystal (see SI). Determining the same for the thin flake, however, requires knowledge of $\alpha_M$ in the in-plane and out-of-plane directions. Lacking $\alpha_M^{\perp}$ for the flake, we can only estimate a range of $\xi_c$=0.4~-~2.5~nm (and $\xi_{ab}$=6~-~14~nm) for $\alpha_M^{\perp}$=6~-~0. In each case, $\xi_c$ is smaller than the sample thickness of 6~nm, but some suppression of orbital pair breaking may still be expected if part of the sample is degraded and the effective superconducting thickness is less than 6~nm.

However, a suppression of orbital pair breaking alone cannot explain our data. If that were the main effect, it would lead to an enhancement of $H_{c2}$ above that of the bulk crystals \cite{VanWeerdenburg2023,Nam2018}. Instead we see a reduction in $H_{c2}$. This suggests it is primarily an enhancement of paramagnetic pair breaking that is responsible for the larger $\alpha_M$, such as might happen due to increased impurity scattering \cite{Khim2010,Fuchs2010}.

As an explanation for our results, we suggest the influence of spin-orbit coupling. We consider the difference between the band structures of monolayer and bulk NbS$_2$, as reported in Ref.~\citenum{Heil2018}. The most apparent difference is in the $p_z$ states at the $\Gamma$ point; in the monolayer, these are pushed below the Fermi energy. However, these states do not play a significant role in the superconductivity of NbS$_2$ \cite{Heil2017}. Instead, we look to the other difference: the effect of spin-orbit coupling. In the monolayer, this causes singlet and triplet pairing channels to mix, leading to a complicated interplay between paramagnetic and orbital pair breaking and a significant $\alpha_M$. In the bulk, on the other hand, we expect only singlet pairing, which may suggest that the Zeeman energy is small compared to the superconducting gap, such that relatively high paramagnetic fields are needed for pair breaking, resulting in $\alpha_M\approx 0$.

Of course, our thin flake is not a monolayer, but the spin-orbit coupling effects seen in monolayers also occur in few-layer samples with odd numbers of layers, as a result of inversion symmetry breaking. As noted earlier, our thin flake sample exhibits clear 2D behavior (see Fig.~\ref{fig2}b) and may thus consist of less than 6~nm of superconducting NbS$_2$ (10 layers), with potentially an odd number of layers. Furthermore, the spin-orbit scattering time in our flake is short ($\tau_{SO}=$186$\pm$78~fs, see SI), as extracted from the WHH fitting. This confirms that spin-orbit coupling is indeed strong and may be responsible for the observed changes in pair breaking. To definitively make this point, however, both theoretical and experimental  work on a series of samples ranging from single layer to several layers will be required. This is beyond the scope of the present work.

We now investigate the Pauli limit of superconductivity, which was reported to be $H_{c2}^{pm}\approx1.86T_c=$10-11~T \cite{Cho2021} in NbS$_2$, a value exceeded by both our bulk and thin flake samples. A possible explanation could lie in the existence of multi-gap superconductivity, a possibility that is also consistent with the observed temperature dependence of the critical field anisotropy \cite{Khim2011a,Gurevich2003} (see supplementary Fig.~S7). Via a different calculation of $H_{c2}^{pm}$ as $H_{c2}^{pm}=\frac{\Delta}{\sqrt{g}\mu_B}$, with the $g$-factor $g=2$ and the superconducting gap $\Delta=1.2$~meV \cite{Diener2011,Guillamon2008,Kacmarcik2010a}, 
 this results in $\mu_0 H_{c2}^{pm}=14.5$~T, below $H_{c2}$ for bulk, but above it for the thin flake. Only the bulk crystal thus appears to violate the Pauli limit.  

In 2D materials, the in-plane upper critical field can be enhanced beyond the Pauli limit due to Ising spin-orbit coupling \cite{DeLaBarrera2018,Huang2022b}, as has been reported for monolayer NbSe$_2$ \cite{Xi2016,Xing2017}. This, however, appears unlikely to occur in bulk NbS$_2$ due to its significant interlayer coupling \cite{Huang2022b}. A violation of the Pauli limit may also occur in conjunction with a superconducting FFLO state, as was previously reported in NbS$_2$ \cite{Cho2021} and NbSe$_2$ \cite{Wan2022}. We note, however, that our data clearly exclude the existence of such a state in NbS$_2$. When entering an FFLO state, the resistive transition should become sharper \cite{Kasahara2020a}, while we observe the opposite (see Fig.~\ref{fig1}(b)). A sharp enhancement of $H_{c2}$ and $H_{irr}$ is also expected at low temperatures and field-angles very close to the in-plane direction \cite{Wosnitza2018}, but is absent here. Finally, there is no sign of a transition between a conventional superconducting and an FFLO state in the magnetostriction.
 
The absence of FFLO in our samples could be attributed to the cleanliness of the crystals. In a clean superconductor, the mean free path $\lambda$ exceeds the coherence length $\xi$, which is a requirement for the presence of an FFLO state. From our complementary Hall effect measurement (see supplementary Fig.~S10), we can estimate that $\lambda\approx$~3-6~nm (see SI for the calculation), suggesting that the superconductivity in our bulk NbS$_2$ is in the dirty regime (i.e. $\lambda<\xi$). However, using a fitting procedure based on the parallel-resistor formalism (see SI and Fig.~S11) \cite{Cooper2009a}, we instead obtain $\lambda\approx$~13~nm. We conclude that our bulk samples are in a crossover between the clean and dirty regimes. We cannot extract $\lambda$ for the thin flake, but we note that even in clean films, surface scattering may destroy the FFLO state \cite{Matsuda2007}.
 
Considering the good agreement between our data and literature reports of RRR \cite{Wen2020,Martino2021,Lian2017,Bag2018}, resistivity \cite{Majumdar2020,Onabe1978,Martino2021}, coherence length \cite{Bi2022,Onabe1978}, carrier density \cite{Molenda1996,Naito1982}, Fermi velocity \cite{Tissen2013}, mobility \cite{Majumdar2020} and the superconducting gap size \cite{Kacmarcik2010a,Diener2011,Guillamon2008}, we conclude that our data present a consistent picture for superconductivity in NbS$_2$. We have thus established the complete $H$-$T$ phase diagram of bulk NbS$_2$ through detailed magnetotransport and magnetostriction measurements and compared it with that of an exfoliated thin flake of NbS$_2$. 

Our main finding is a strongly enhanced Maki parameter with reduced $H_{c2}$ in a thin flake compared to bulk NbS$_2$ crystals, suggesting a suppression of orbital pair breaking coupled with an enhancement of paramagnetic pair breaking. A possible explanation is given by spin-orbit coupling that affects few-layer NbS$_2$ flakes with odd layer numbers. Considering the unexpected two-dimensional behavior, this may be applicable to the thin flake sample we have studied. In future work, a systematic study of NbS$_2$ flakes with even and odd numbers of layers is required to understand the influence of spin-orbit coupling on the superconducting pair breaking in this material. By studying a wide range of thicknesses, it will be possible to identify the crossover regime where paramagnetic pair breaking begins to play a role. One may then think about purposely controlling the dominant pair breaking mechanism, enabling a further understanding of these systems and greater control over their superconductivity.

		\begin{acknowledgments}
The authors acknowledge fruitful discussions with M. Katsnelson. This work was supported by HFML-RU/NWO-I, member of the European Magnetic Field Laboratory (EMFL). This publication is part of the project TOPCORE (OCENW.GROOT.2019.048) of the research program NWO – GROOT which is financed by the Dutch Research Council (NWO). 
		\end{acknowledgments}

	\end{document}